\documentclass[11pt, twoside]{article}

\usepackage{newpasp,epsf}

\begin{document}

\title{Anatomy of giant spiral galaxies}
\author{E. Pignatelli}
\affil{SISSA/ISAS, via Beirut 4, I-34014 Trieste, Italy}
\author{G. Galletta}
\affil{Dipartimento di Astronomia, Universit\`a di Padova}


\section{Introduction}

The progress made in last years in numerical modeling
allows to study the internal structure of galaxies with
great detail. This is of particular interest for stellar
systems composed by several components, such as spiral
galaxies. In the recent years, we produced a
self-consistent, 3-D model of disk galaxy made by an
exponential disk and a $r^{1/4}$ bulge (Pignatelli \&
Galletta, 1999).

The model fit simultaneously the photometric data
(surface brightness, ellipticities and PAs of isophote
major axis) and the kinematic data (rotation curves and
velocity dispersion curves along various PAs). In such a
way, the structure of the luminous matter derived from
the photometric analysis is compared with the total
matter deduced from the dynamical data and the presence
of dark matter may be studied. 

To test the model, we selected five galaxies with
morphological types spanning from S0/a to Sc and absolute
magnitudes from -20.6 to -22.5. They have been selected
from HI observations looking for objects with a line
width $W_{20} > 350$ km/sec. 

\begin{figure}[h]
\plottwo{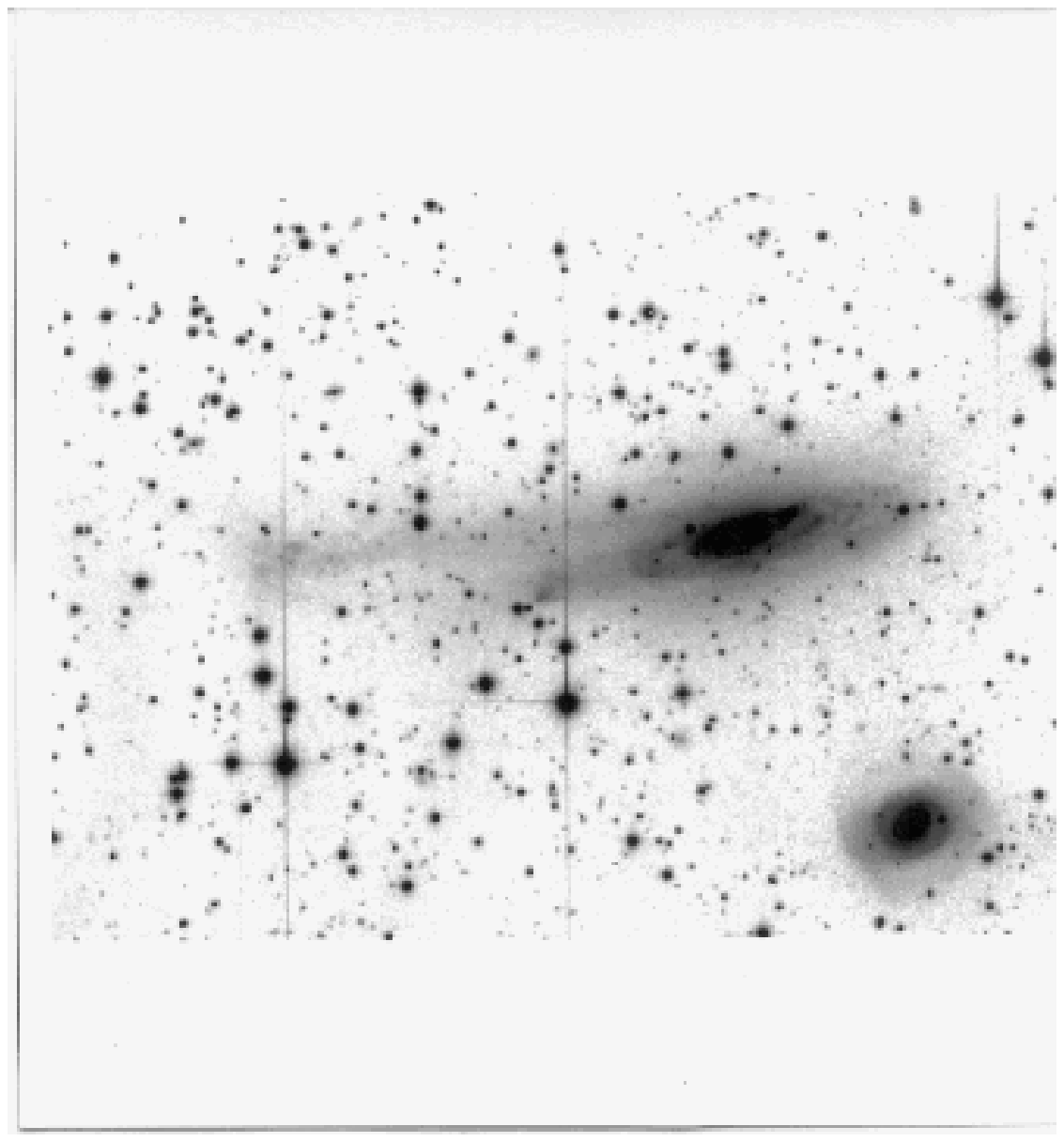}{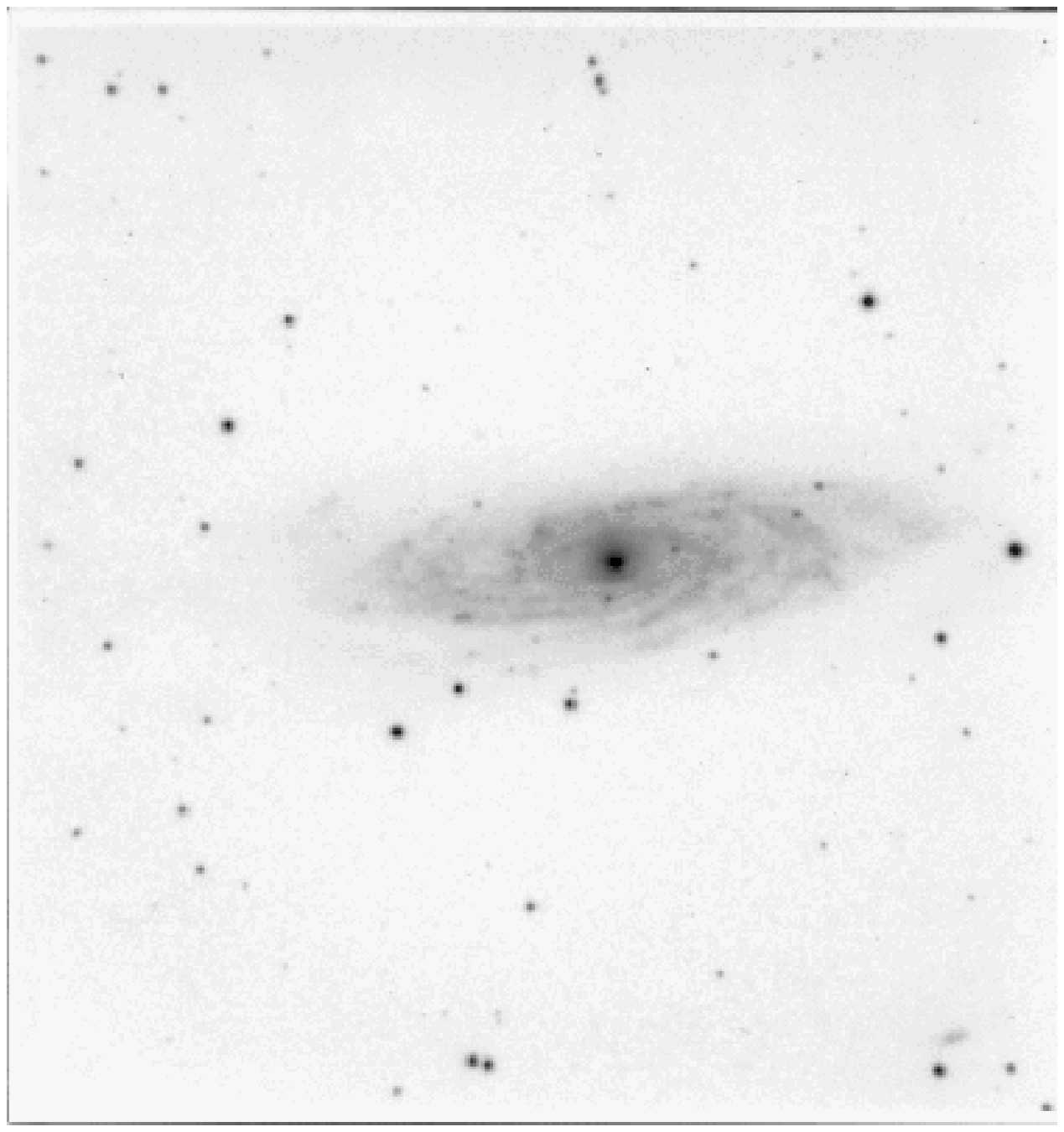}
\caption{B images of two galaxies of the sample: the SBcd
galaxy NGC 3263 and the Sbc galaxy NGC 6925.}
\end{figure}

For all these galaxies we obtained images at B and I
photometric wave bands using the ESO telescopes at La
Silla. The used telescopes were the Danish 1.5m, the
NTT+EMMI and the ESO-MPI 2.2m+EFOSC2. The spectra have
been collected with the spectrographs Boller\& Chivens
and EFOSC2 of the 2.2m ESO-MPI telescope.

We used the model to evaluate the intrinsic properties of
these galaxies, such as the disk/bulge mass ratio and the
scale length of the galaxy components. 

\vspace*{-0.4cm}
\section{The results from the model}

The galaxies can be described by superposition of
different components. For each component, we separately
assume that: 
(a) the density distribution is oblate, without
strongly triaxial structures;
(b) the isodensity surfaces are similar concentric
spheroids;
(c) the surface density profile follows a simple
$R^{1/4}$ or exponential law;
(d) the velocity distribution is locally Gaussian;
(e) the velocity dispersion is isotropic
$\sigma_r=\sigma_\theta=\sigma_z$;

The model has $4n+1$ free parameters, where $n$ is the
number of adopted components: namely the luminosity
$L_{tot}$, the effective radius $R_e$, the
mass-luminosity ratio $M/L$ and the flattening $b/a$ of
each component plus the inclination angle $i$ of the
galaxy. 

Photometry can be used to constrain all these parameters
except the M/L ratio, which must be derived by
kinematics. In Fig. 2 we show an example of the
application of the model to the rotation curve of
NGC~6925: in this case, we found that the stellar
velocity can be fitted by a model without dark matter
with a $M/L=3.9$ solar units and a total mass of $2.7
\cdot 10^{11} M_\odot$. 

\begin{figure}[h]
\vspace*{4cm}
\includegraphics{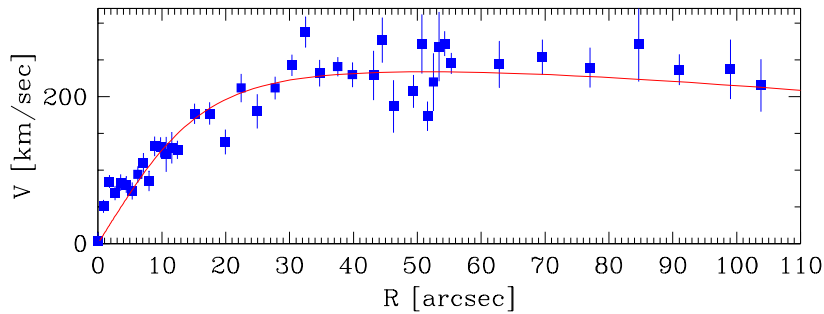}
\caption{Stellar rotation velocities for NGC 6925,
compared with the results of the dynamical model (see text).}
\end{figure}

\noindent 
Since these galaxies have high value for the HI line
width $W_{20}$, it has been suggested (Buson et al.,
1991) that they could have in general an unusually high
content of dark matter in the inner regions or, perhaps,
an unusual stellar population. None of this hypothesis is
confirmed by our  analysis.

\end{document}